\documentclass[aps,prd,showpacs,showkeys,preprint]{revtex4}
\usepackage[british]{babel}

\usepackage{amsmath}
\usepackage{amssymb}

\usepackage{bbm}

\DeclareFontFamily{OT1}{rsfs}{}
\DeclareFontShape{OT1}{rsfs}{m}{n}{ <-7> rsfs5 <7-10> rsfs7 <10->rsfs10}{} 
\DeclareMathAlphabet{\mycal}{OT1}{rsfs}{m}{n}
\newcommand{\lan}{{\mycal L}}

\newcommand{\p}{\partial}

\newcommand{\vp}{\mathbf{p}}
\newcommand{\vo}{\mathbf{0}}

\newcommand{\dual}[1]{\overset{{}^{{}^{\boldsymbol{\neg}}}}{\smash[t]{#1}}}

\begin{document}

\title{The Einstein-Elko system -- Can dark matter drive inflation?}

\author{C. G. B\"ohmer}
\email{christian.boehmer@port.ac.uk}
\affiliation{Institute of Cosmology and Gravitation,
             University of Portsmouth, Portsmouth, PO1 2EG, UK}
\affiliation{ASGBG/CIU, Department of Mathematics, Apartado Postal C-600, 
             University of Zacatecas (UAZ), Zacatecas, Zac 98060, Mexico}

\date{16 January 2007}

\begin{abstract}
Recently, a spin one half matter field with mass dimension one was discovered,
called Elko spinors. The present work shows how to introduce these 
fields into a curved spacetime by the standard covariantisation scheme. After 
formulating the coupled Einstein-Elko field equations, the spacetime is assumed 
to be homogeneous and isotropic in order to simplify the resulting field equations. 
Analytical ghost Elko solutions are constructed which have vanishing 
energy-momentum tensor without and with cosmological constant. The cosmological Elko 
theory is finally related to the standard scalar field theory with self interaction 
that gives rise to inflation and it is pointed out that the Elko spinors are 
not only prime dark matter candidates but also prime candidates for inflation.
\end{abstract}

\pacs{04.40.-b, 95.35.+d, 98.80.-k}

\keywords{Elko spinors, non-standard spinors, inflation, dark matter}

\maketitle

\section{Introduction}

Recently and quite unexpectedly a spin one half matter field with mass dimension 
one was discovered in the two papers~\cite{jcap} and~\cite{prd} by 
Ahluwalia-Khalilova and Grumiller. These non-standard Wigner class spinors are 
based on the eigenspinors of the charge conjugation operator. The resulting field 
theory is non-local, having the unusual property $(CPT)^2=-\mathbbm{1}$. They 
belong to a wider class of so-called flagpole spinors, see~\cite{daRocha:2005ti} 
for the Lounesto spinor field classification of Elko spinors, they are of class 5.
It should be emphasised that class 4 spinors have not been studied so far,
therefore it might be interesting to initiate their analysis, also as possible
dark matter candidates.

The dominant interactions~\cite{jcap,prd} 
of this field with standard matter only take place through the Higgs doublet 
or with gravity. These properties make the Elko spinors a prime candidate for 
dark matter which is based on first principles. This very attractive feature 
motivates further research on these fields.

The next logical step is the introduction of these spinor fields into an
arbritrary spacetime and the analysis of the coupled Einstein-Elko field equations,
which is the aim of the present work. Elko spinors are non-local, but
representations of the Poincar\'e group and therefore the covariant derivative
can easily be extended. This enables one to construct the Lagrangian and hence 
the coupled Einstein-Elko system can be studied.

The paper is organised as follows. In Section~\ref{sec:d} the covariant derivative
when acting on an Elko spinor is computed and its general relativistic Lagrangian 
is constructed. In Section~\ref{sec:cos} the spacetime is assumed to be homogeneous
and isotropic, and cosmological Elko spinors satisfying the cosmological principle
are derived. After this kinematical discussion, the coupled Einstein-Elko field
equations are studied in Section~\ref{sec:dyn}. The Elko field theory as a 
prime candidate to drive inflation is presented in Section~\ref{inflation}. 
Conclusions and an outlook are given in the final Section~\ref{sec:end}.

\section{Dirac and Elko spinors in spacetime}
\label{sec:d}

In this section the covariant derivative, when acting on an Elko spinor
is calculated. To do so, let us recall the corresponding procedure
in case of Dirac spinors. In order to introduce spinor fields let 
us define the tetrad field $e^{\mu}_{a}$ such that
\begin{align}
      e^{\mu}_{a} e^{\nu}_{b} \eta^{ab} = g^{\mu\nu} \,,
      \label{eq:d0}
\end{align}
where $(a,b,\ldots)$ can take the values $(0,1,\ldots)$ and
$(\mu,\nu,\ldots)$ take the values $(t,\ldots)$, the names of the 
coordinates. $(a,b,\ldots)$ are the anholonomic indices, whereas
$(\mu,\nu,\ldots)$ are the holonomic indices. Throughout the paper
we use the signature $(+,-,-,-)$.
 
Let a Dirac spinor in Weyl representation be given by
\begin{align}
      \psi = \begin{pmatrix} \phi_R \\ \phi_L \end{pmatrix} \,,
      \label{eq:d1}
\end{align} 
and the $\gamma$-matrices in chiral form
\begin{align}
      \gamma^{0} = \begin{pmatrix} \mathbb{O} & \mathbbm{1} \\
                 \mathbbm{1} & \mathbb{O} \end{pmatrix} \,, \quad
      \gamma^{i} = \begin{pmatrix} \mathbb{O} & -\sigma^i \\
                 \sigma^i & \mathbb{O} \end{pmatrix} \,, \quad
      \gamma^{5} = \begin{pmatrix} \mathbbm{1} & \mathbb{O} \\
                 \mathbb{O} & -\mathbbm{1} \end{pmatrix} \,,
      \label{eq:d2}
\end{align}
with $\gamma^{5} = i\gamma^{0}\gamma^{1}\gamma^{2}\gamma^{3}$.
For sake of comparison, the conventions and notation of~\cite{jcap}
are adopted. The $\gamma$-matrices satisfy
\begin{align}
      \gamma^{(a} \gamma^{b)} = \eta^{ab} \,, \qquad
      \gamma^{(\mu} \gamma^{\nu)} = g^{\mu\nu} \,,
      \label{eq:d2a}
\end{align}
where $\gamma^{\mu} = e^{\mu}_{a} \gamma^a$. The covariant derivative of 
a Dirac spinor is defined by
\begin{align}
      \nabla_a \psi = \p_a \psi -\Gamma_{a} \psi \,,
      \label{eq:d3}
\end{align}
where $\Gamma_{a}=\Gamma_{abc}f^{bc}$ and $f^{bc}$ denotes the generators 
of the Lorentz group, $f^{bc}=\frac{1}{4}[\gamma^{b},\gamma^{c}]$, so that 
the covariant derivative~(\ref{eq:d3}) becomes
\begin{align}
      \nabla_a \psi = \p_a \psi - \Gamma_a \psi \,, \qquad
      \Gamma_{a} = \frac{1}{4} \Gamma_{abc} \gamma^b \gamma^c \,.
      \label{eq:d4}
\end{align} 
The covariant derivative when acting on a vector $A^{\mu}$ is given by
\begin{align}
      \nabla_{\mu} A^{\nu} = \p_{\mu} A^{\nu} + 
      \Gamma^{\nu}_{\mu\kappa} A^{\kappa} \,.
      \label{eq:d4a}
\end{align}
The Christoffel symbols $\Gamma$ in anholonomic $(\Gamma^{a}_{bc})$ and 
holonomic $\Gamma_{\mu\nu}^{\kappa}$ form are related by
\begin{align}
      \Gamma^{c}_{ab} = e^{\mu}_{a} e^{\nu}_{b} \nabla_{\mu} e_{\nu}^{c} =
      e^{\mu}_{a} e^{\nu}_{b} e^{c}_{\kappa} \Gamma_{\mu\nu}^{\kappa}
      - e^{\mu}_{a} e^{\nu}_{b} \p_{\mu }e^{c}_{\nu} \,,
      \label{eq:d4b} \\
      \Gamma_{abc} = g_{cd} \Gamma_{ab}^{d}\,, \qquad \Gamma_{a(bc)}=0 \,.
      \label{eq:d4c}
\end{align} 
Alternatively, the covariant derivative~(\ref{eq:d4}) can be written
with a holonomic index
\begin{align}
      \nabla_{\mu} \psi = \p_{\mu} \psi - \Gamma_{\mu} \psi \,, \qquad
      \Gamma_{\mu} = \frac{1}{4} e_{\nu}^{b}(\nabla_{\mu} e^{\nu}_c) \gamma_b \gamma^c \,,
      \label{eq:d4d}
\end{align}
where equation~(\ref{eq:d4b}) was taken into account.

With the Dirac adjoint given by $\bar{\psi}=\psi^{\dagger} \gamma^0$, 
the covariant derivative acting on the scalar $\bar{\psi}\psi$ is known. 
This defines the covariant derivative when acting on $\bar{\psi}$ to be 
\begin{align}
      \nabla_a \bar{\psi} = \p_a \bar{\psi} + \bar{\psi} \Gamma_a \,.
      \label{eq:d4e}
\end{align}
The following relations are quite useful for what follows
\begin{align}
      \gamma^0 \gamma^0 = \mathbbm{1}\,, \quad \gamma^i \gamma^i = -\mathbbm{1}\,,
      \quad \gamma^0 \gamma^i = \begin{pmatrix} \sigma^i & \mathbb{O} \\ 
                                \mathbb{O} & -\sigma^i \end{pmatrix}\,, \\[1ex] 
      \gamma^i \gamma^j = - \begin{pmatrix} \sigma^i \sigma^j & \mathbb{O} \\
                            \mathbb{O} & \sigma^i \sigma^j \end{pmatrix}
                        = -i\varepsilon^{ijk}\begin{pmatrix} \sigma_k & \mathbb{O} \\
                            \mathbb{O} & \sigma_k \end{pmatrix}\,.
      \label{eq:d6}
\end{align}
From equations~(\ref{eq:d1}) and~(\ref{eq:d2}) one can define how the
covariant derivative acts on the right- and left-handed two-spinor of
the Dirac four-spinor. Take into account that the second term of 
equation~(\ref{eq:d4}) can be written as follows
\begin{align}
      \Gamma_{abc} \gamma^b \gamma^c&= \Gamma_{a0i} \gamma^0 \gamma^i +
      \Gamma_{ai0} \gamma^i \gamma^0 + \Gamma_{aij} \gamma^i \gamma^j 
      \nonumber \\
      &= 2 \Gamma_{a0i} \gamma^0 \gamma^i + \Gamma_{aij} \gamma^i \gamma^j
      \nonumber \\
      &= 2 \Gamma_{a0i} \begin{pmatrix} \sigma^i & \mathbb{O} \\ 
                        \mathbb{O} & -\sigma^i \end{pmatrix}
      -i \Gamma_{aij} \varepsilon^{ijk}\begin{pmatrix} \sigma_k & \mathbb{O} \\
                      \mathbb{O} & \sigma_k \end{pmatrix}
      \label{eq:d7}
\end{align}
Therefore, the covariant derivative~(\ref{eq:d4}) can be written as two 
two-spinor equations
\begin{align}
      \nabla_a \phi_R = \p_a \phi_R + \bigl(+2\Gamma_{a0i} \sigma^i 
      - i \Gamma_{aij} \varepsilon^{ijk} \sigma_k \bigr) \phi_R \,,
      \label{eq:right} \\
      \nabla_a \phi_L = \p_a \phi_L + \bigl(-2\Gamma_{a0i} \sigma^i 
      - i \Gamma_{aij} \varepsilon^{ijk} \sigma_k \bigr) \phi_L \,.
      \label{eq:left}
\end{align}
Having worked out the action of the covariant derivative on the left- and
right-handed spinors respectively, one can now turn to the computation of
the covariant derivative, when acting on an Elko spinor. Such an Elko spinor 
is defined by~\cite{jcap,prd}
\begin{align}
      \lambda = \begin{pmatrix} \pm \sigma_2 \phi^{\ast}_{L} \\
                \phi_L \end{pmatrix} \,,
      \label{eq:d5}
\end{align}
where $\phi^{\ast}_{L}$ denotes complex conjugation of $\phi_L$.
The upper sign is for the self conjugate spinor and the lower sign
for the anti self conjugate spinor with respect to the charge
conjugation operator. One could have equally well started with the 
right-handed two-spinor $\phi_{R}$, however both constructs carry
the same physical content. All details regarding the field theory of
Elko spinors can be found in Ref.~\cite{jcap}, with a condensed version
published in~\cite{prd}.

One can now compute the covariant derivative, when acting on $\lambda$,
by firstly computing the covariant derivative of $\phi_L$, and secondly
by complex conjugation and multiplication by $\sigma_2$ from the left of
the resulting equation.

Start by complex conjugating the covariant derivative of the 
left-handed Dirac spinor~(\ref{eq:left})
\begin{align}
      \nabla_a \phi^{\ast}_L = \p_a \phi^{\ast}_L + 
      \bigl(-2\Gamma_{a0i} (\sigma^i)^{\ast} 
      + i \Gamma_{aij} \varepsilon^{ijk} (\sigma_k)^{\ast} \bigr) \phi^{\ast}_L \,,
      \label{eq:left1}
\end{align} 
and multiply the latter equation by $\sigma^2$ from the left
\begin{align}
      \nabla_a \sigma^2 \phi^{\ast}_L &= \p_a \sigma^2 \phi^{\ast}_L + 
      \bigl(-2\Gamma_{a0i} \sigma^2 (\sigma^i)^{\ast} + i \Gamma_{aij} \varepsilon^{ijk} 
      \sigma^2 (\sigma_k)^{\ast} \bigr) \phi^{\ast}_L \,.
      \label{eq:left2}
\end{align}
For the second term of the right side we obtain
\begin{align}
      \text{if}\ i&=2:\quad -2\Gamma_{a02} \sigma^2 (\sigma^2)^{\ast} \phi^{\ast}_L = 
                       2\Gamma_{a02} \sigma^2 \sigma^2 \phi^{\ast}_L \,, \\
      \text{if}\ i&\neq2:\quad -2\Gamma_{a0i} \sigma^2 (\sigma^i)^{\ast} \phi^{\ast}_L =
                         -2\Gamma_{a0i} \sigma^2 \sigma^i \phi^{\ast}_L =
                          2\Gamma_{a0i} \sigma^i \sigma^2 \phi^{\ast}_L \,,
\end{align}
and for the third term
\begin{align}
      \text{if}\ k&=2:\quad i \Gamma_{aij} \varepsilon^{ij2} \sigma^2 
                      (\sigma_2)^{\ast} \phi^{\ast}_L = 
                           -i \Gamma_{aij} \varepsilon^{ijk} 
                       \sigma^2 \sigma_2 \phi^{\ast}_L \,, \\
      \text{if}\ k&\neq2:\quad i \Gamma_{aij} \varepsilon^{ijk} \sigma^2 
                      (\sigma^k)^{\ast} \phi^{\ast}_L =
                            i \Gamma_{aij} \varepsilon^{ijk} \sigma^2 
                       \sigma_k \phi^{\ast}_L =
                            -i \Gamma_{aij} \varepsilon^{ijk} \sigma_k 
                       \sigma^2 \phi^{\ast}_L \,.
      \label{eq:left3}
\end{align}
Combining the equations~(\ref{eq:left2})--(\ref{eq:left3}) yields the
covariant derivative, when acting on $\pm \sigma^2 \phi^{\ast}_L$, to be
\begin{align}
       \nabla_a (\pm \sigma^2 \phi^{\ast}_L) = \p_a (\pm \sigma^2 \phi^{\ast}_L) + 
       \bigl(+2\Gamma_{a0i} \sigma^i -i \Gamma_{aij} \varepsilon^{ijk} 
       \sigma_k \bigr) (\pm \sigma^2 \phi^{\ast}_L) \,,
       \label{eq:left4}
\end{align}
which has exactly the same formal structure as the covariant derivative,
when acting on the right-handed spinor~(\ref{eq:right}).

Hence, the covariant derivative on an Elko spinor $\lambda$ can simply
be written in the usual form
\begin{align}
      \nabla_a \lambda = \p_a \lambda - \Gamma_{a} \lambda \,, \qquad
      \Gamma_{a} = \frac{1}{4} \Gamma_{abc} \gamma^b \gamma^c \lambda \,.
      \label{eq:d8}
\end{align}

Before defining an Elko dual, the double helicity structure of the spinor
$\lambda$ (\ref{eq:d5}) should be emphasised (with respect to the standard
Dirac adjoint $\bar{\psi} = \psi^{\dagger}\gamma^0$ the Elkos have an imaginary
bi-orthogonal norm~\cite{jcap}).

The helicities of $\phi_L$ and $\sigma^2 \phi^{\ast}_L$ are opposite~\cite{jcap}, 
to distinguish the two possible helicity configurations we write
\begin{align}
      \lambda_{\{-,+\}} = \begin{pmatrix} \pm \sigma_2 \phi^{+}_{L}{}^{\ast} \\
                \phi^{+}_L \end{pmatrix} \,, \qquad
      \lambda_{\{+,-\}} = \begin{pmatrix} \pm \sigma_2 \phi^{-}_{L}{}^{\ast} \\
                \phi^{-}_L \end{pmatrix} \,.
      \label{eq:d9}
\end{align}
The first entry of the helicity subscript $\{-,+\}$ refers to the upper two-spinor
and the second the lower two-spinor. Let us henceforth denote the helicity
subscript by the indices $u,v,\ldots$ and define the Elko dual by
\begin{align}
       \dual{\lambda}_u = i\,\varepsilon_u^v \lambda_v^{\dagger} \gamma^0 \,,
       \label{eq:d10}
\end{align}
with the skew-symmetric symbol 
$\varepsilon_{\{+,-\}}^{\{-,+\}}=-1=-\varepsilon_{\{-,+\}}^{\{+,-\}}$.
With the dual defined above one finds (by construction)
\begin{align}
      \dual{\lambda}_u(\vp) \lambda_v(\vp) = \pm 2m \delta_{uv} \,,
      \label{eq:d11}
\end{align}
where $\vp$ denotes the momentum.

Since the covariant derivative $\nabla_a$ obeys the Leibniz rule and its action
on the scalar $\dual{\lambda}\lambda$ is known, one can construct the covariant
derivative when acting on $\dual{\lambda}$ (like in the Dirac case)
\begin{align}
      \nabla_a \dual{\lambda} = \p_a \dual{\lambda} + 
      \dual{\lambda} \Gamma_{a} \,.
      \label{eq:d12}
\end{align}
It is not very surprising that one obtains the same covariant derivative for
the Dirac and the Elko spinors. This is due to the fact that the Lorentz
generators enter the definition of the covariant derivative, and their
structure is unaffected by the Elko spinors.

After deriving the covariant derivative on an Elko spinor, one can formulate the 
Elko Lagrangian. It is tempting to use a kinetic term of the form 
$g^{\mu\nu}\nabla_{\mu} \dual{\lambda} \nabla_{\nu} \lambda$. Note that this
term if varied with respect to the metric tensor $g^{\mu\nu}$ yields 
the term $\nabla_{\mu} \dual{\lambda} \nabla_{\nu} \lambda $ which is 
not necessarily symmetric due to the fact that $\lambda$ is a spinor. However,
adding the (Elko) conjugate leads to a symmetrised kinetic term. Therefore, 
the following action is obtained
\begin{multline}
      S_{\rm elko} = \int \sqrt{-g}\, \lan_{\rm elko}\, d^4 x 
      = \int L_{\rm elko }\, d^4 x \\=
      \frac{1}{2} \int \sqrt{-g} \Bigl( \frac{1}{2} g^{\mu\nu}
      \bigl(\nabla_{\mu} \dual{\lambda} \nabla_{\nu} \lambda
      + \nabla_{\nu} \dual{\lambda} \nabla_{\mu} \lambda \bigr)
      -m^2 \dual{\lambda} \lambda + \alpha 
      [\dual{\lambda} \lambda]^2 \Bigr) d^4 x\,,
      \label{eq:d14}
\end{multline}
where $\lan_{\rm elko}$ denotes the Lagrangian density, 
$L_{\rm elko} = \sqrt{-g}\, \lan_{\rm elko}$ the Elko Lagrangian,
and lastly $S_{\rm elko}$ denotes the Elko spinor action. With $g$ 
we denote the determinant of the metric.

Before studying the coupled cosmological Einstein-Elko field equations,
let us note that the helicity up and down spinors cannot be chosen
independently. In the zero momentum rest frame the spinors are explicitely
(see~\cite{prd}) given by
\begin{align}
      \phi^{+}_L(\vo) &= \sqrt{m} \begin{pmatrix} \cos(\theta/2) e^{-i\phi/2} \\
                   \sin(\theta/2) e^{i\phi/2}\end{pmatrix} \,, 
      \nonumber \\[1ex]
      \phi^{-}_L(\vo) &= \sqrt{m} \begin{pmatrix} \sin(\theta/2) e^{-i\phi/2} \\
                   -\cos(\theta/2) e^{i\phi/2}\end{pmatrix} \,,
      \label{eq:d15}
\end{align}
where $\phi^{\pm}_L$ are helicity eigenstates that satisfy 
$\mathbf{\sigma}\cdot\hat{\vp}\,\phi^{\pm}_L(\vo)=\pm\phi^{\pm}_L(\vo)$, and
where $\hat{\vp}=(\sin\theta\cos\phi, \sin\theta\sin\phi, \cos\theta)$
is the unit momentum. In order to simplify the following analysis as much as 
possible we will, unless otherwise stated, take $\hat{\vp}=\mathbf{z}=(0,0,1)$, 
i.e.~$\theta=0$ and $\phi=0$. This choice yields the following
rest frame spinors 
\begin{align}
      \phi^{+}_L(\vo) = \sqrt{m} \begin{pmatrix} 1 \\ 0 \end{pmatrix} \,, \qquad
      \phi^{-}_L(\vo) = \sqrt{m} \begin{pmatrix} 0 \\-1 \end{pmatrix} \,,
      \label{eq:d16}
\end{align}
which can be boosted to arbitrary momentum following~\cite{jcap,prd}. The
Elko spinors constructed from these building blocks, with the dual defined
by~(\ref{eq:d10}) satisfy the above mentioned orthogonality relation~(\ref{eq:d11})
and also satisfy the completeness relation
\begin{align}
      \sum_u \Bigl( \lambda^S_u(\vp) \dual{\lambda}^S_u(\vp) - 
      \lambda^A_u(\vp) \dual{\lambda}^A_v(\vp) \Bigr) = 2m\, \mathbbm{1} \,,
      \label{eq:d17}
\end{align}
where the superscripts $^S$ and $^A$ denote the self and anti self conjugate
spinors.

\section{Cosmological kinematics} 
\label{sec:cos}

In this section the coupled Einstein-Elko system is discussed assuming
that the spacetime is homogeneous and isotropic. Solutions of the coupled
Einstein-Elko system will be discussed in the next section. Here the
focus is more on the kinematical aspects.

The metric in Friedman-Robertson-Walker (FRW) form is given by
\begin{align}
      ds^2 = dt^2 - \Bigl(\frac{a(t)}{1-\frac{k}{4}r^2}\Bigr)^2 
      \bigl( dx^2 + dy^2 + dz^2 \bigr) \,,
      \label{eq:cos1}
\end{align}
where $a(t)$ is the expansion parameter, that defines the Hubble
parameter by $H(t)=\dot{a}(t)/a(t)$. Therefore we furthermore assume
that the Elko spinors also only depends on the time coordinate $t$. For
further simplification, and in agreement with the current observations
in cosmology~\cite{Riess:1998cb,Perlmutter:1998}, the constant time slices 
are chosen to be flat, $k=0$, so that the FRW 
metric~(\ref{eq:cos1}) becomes
\begin{align}
      ds^2 = dt^2 - a(t)^2 \bigl( dx^2 + dy^2 + dz^2 \bigr) \,.
      \label{eq:cos1a}
\end{align}

This metric yields the following non-vanishing holonomic Christoffel symbols
$\Gamma_{\mu\nu}^{\kappa}$ 
\begin{align}
      \Gamma_{tx}^{x} =  \Gamma_{ty}^{y} =  \Gamma_{tz}^{z} &= \frac{\dot{a}}{a} \,, 
      \nonumber \\
      \Gamma_{xx}^{t} =  \Gamma_{yy}^{t} =  \Gamma_{zz}^{t} &= a\dot{a} \,,
      \label{eq:cos2}
\end{align}
where the dot denotes differentiation with respect to $t$.
Consequently, the non-vanishing Christoffel symbols $\Gamma_{\mu}$ for 
the covariant derivative on spinors~(see~(\ref{eq:d4d})) are given by  
\begin{align}
      \Gamma_x = \dot{a} \frac{1}{4}
      \Bigl( \gamma^0 \gamma^1 - \gamma^1 \gamma^0 \Bigr) 
      = \dot{a} f^{01} \,, \nonumber \\
      \Gamma_y = \dot{a} \frac{1}{4}
      \Bigl( \gamma^0 \gamma^2 - \gamma^2 \gamma^0 \Bigr)  
      = \dot{a} f^{02} \,, \nonumber \\
      \Gamma_z = \dot{a} \frac{1}{4}
      \Bigl( \gamma^0 \gamma^3 - \gamma^3 \gamma^0 \Bigr)
      = \dot{a} f^{03} \,,
      \label{eq:cos3}
\end{align}
where $f^{ab}$ are the generator of the Lorentz group (see after~(\ref{eq:d3})).

The Lagrangian for the Elko spinors and the massive complex scalar field
Lagrangian are formally very similar with the only difference that the one
is a spinor while the other is a scalar field. It is therefore quite useful
to shortly review the energy-momentum tensor and the matter field equations
for complex scalars.

\subsection{Complex scalars}

Let us start the kinematical discussion with a massive complex scalar 
field $\phi$, whose Lagrangian reads
\begin{align}
      L_{\bar{\phi}\phi} = \frac{1}{2} \sqrt{-g} \bigl( g^{\mu\nu}
      \nabla_{\mu} \bar{\phi} \nabla_{\nu} \phi 
      -m^2_{\phi} \bar{\phi} \phi \bigr) \,,
      \label{eq:cs1}
\end{align}
where we omitted the self-interaction term $\alpha[\bar{\phi}\phi]^2$ to keep
the following as simple as possible.

Although the action of a complex scalar field is formally equal to the Elko 
action~(\ref{eq:d14}) they are very different because of the spinor 
structure of $\lambda$. The massive complex scalar's energy-momentum 
tensor is given by 
\begin{align}
      T_{\mu\nu} = \frac{2}{\sqrt{-g}}\frac{\delta L_{\bar{\phi}\phi}}
      {\delta g^{\mu\nu}} = \nabla_{\mu} \bar{\phi} \nabla_{\nu} \phi 
      - g_{\mu\nu}\lan_{\bar{\phi}\phi} \,.
      \label{eq:cs2}
\end{align}
By taking the FRW metric~(\ref{eq:cos1a}) into account we explicitely get
for the Lagrangian density of massive complex scalars
\begin{align}
      \lan_{\bar{\phi}\phi} = \frac{1}{2} \bigl( \p_t \bar{\phi} \p_t \phi
      -m^2_{\phi} \bar{\phi} \phi \bigr) \,,
      \label{eq:cs3}
\end{align} 
which can be substituted into~(\ref{eq:cs2}) to yield the explicit form
of the energy-momentum tensor
\begin{align}
      T_{tt} &= \p_t \bar{\phi}\, \p_t \phi - \frac{1}{2} 
      \bigl( \p_t \bar{\phi} \p_t \phi - m^2_{\phi} \bar{\phi} \phi \bigr) 
      = \frac{1}{2} \bigl(\p_t \bar{\phi} \p_t \phi 
      + m^2_{\phi} \bar{\phi} \phi  \bigr) \,, 
      \label{eq:cs4} \\
      T_{xx} &= T_{yy} = T_{zz} = \frac{a^2}{2} 
      \bigl( \p_t \bar{\phi} \p_t \phi - m^2_{\phi} \bar{\phi} \phi \bigr)\,.
      \label{eq:cs5}
\end{align}
The off-diagonal terms all identically vanish, so that the requirement
of homogeneity and isotropy does {\em not} imply further matter restrictions. 
It is this point that will change significantly, when the energy-momentum
tensor for the Elko spinors is derived.

Variation of the complex scalar field action~(\ref{eq:cs1}) with respect 
to $\bar{\phi}$ and $\phi$ leads to the following equations of motions
\begin{align}
      \Box \phi + m^2_{\phi} \phi = 0 \,, \qquad
      \Box \bar{\phi} + m^2_{\phi} \bar{\phi} = 0 \,,
      \label{eq:cs6}
\end{align}
where the operator $\Box\phi$ for scalars reads
\begin{align}
      \Box\phi = \p_{tt}\phi + 3\Bigl(\frac{\dot{a}}{a}\Bigr) \p_t \phi \,.
      \label{eq:cs7}
\end{align}
{\bf A cautionary note:~}(see~\cite{Deser:2004yh}) Variation of~(\ref{eq:cs3}) 
with respect to $\bar{\phi}$ does not lead to the correct equations of 
motion~(\ref{eq:cs6}), instead one would arrive at
\begin{align}
      \p_{tt} \phi + m^2_{\phi} \phi = 0 \,,
      \label{eq:cs8}   
\end{align} 
together with the conjugate equation, which does not take into account 
the second term of the $\Box\phi$ operator according to~(\ref{eq:cs7}).

The reason why the equations of motion of the reduced theory are not
equivalent with the equations of motion of the original theory, when
the latter is restricted to homogeneity and isotropy lies in the fact
that the ``principle of symmetric criticality''~\cite{Palais:1979}
cannot be applied in that case (opposed to spherical symmetry where
the equations of motion are equivalent). In that respect the reader is
referred to~\cite{Fels:2001rv,Deser:2003up}.

\subsection{Elko spinors in cosmology}
\label{ssec:elko}

The calculations of the previous subsection are now repeated for the Elko spinors. 
It turns out that things change considerably due to the spinor nature of Elkos.
Variation of its Lagrangian~(\ref{eq:d14}) with respect to the metric
tensor $g^{\mu\nu}$ yield the following energy-momentum tensor
\begin{align}
      T_{\mu\nu} = \frac{1}{2}\bigl(\nabla_{\mu} \dual{\lambda} 
      \nabla_{\nu} \lambda + \nabla_{\nu} \dual{\lambda} 
      \nabla_{\mu} \lambda \bigr) - g_{\mu\nu} \lan_{\rm elko} \,.
      \label{eq:cos4}
\end{align}
Start with the explicit form of the Lagrangian density $\lan_{\rm elko}$ by taking
the FRW metric~(\ref{eq:cos1a}) into account. The kinetic term becomes
\begin{align}
      g^{\mu\nu} \nabla_{\mu} \dual{\lambda} \nabla_{\nu} \lambda 
      &= g^{\mu\nu} \bigl(\p_{\mu} \dual{\lambda} + \dual{\lambda} \Gamma_{\mu}\bigr)
      \bigl(\p_{\mu} \lambda - \Gamma_{\mu} \lambda\bigr) 
      \nonumber \\
      &= g^{\mu\nu} \p_{\mu} \dual{\lambda} \p_{\mu} \lambda -
      \dual{\lambda} g^{\mu\nu} \Gamma_{\mu} \Gamma_{\nu} \lambda 
      \label{eq:cos4a}
\end{align}
where we took into account that $\Gamma_t=0$ and $\p_{x,y,z}\lambda=0$,
and therefore find
\begin{align} 
      g^{\mu\nu} \nabla_{\mu} \dual{\lambda} \nabla_{\nu} \lambda = 
      \p_t \dual{\lambda} \p_t \lambda + \frac{3}{4} 
      \Bigl(\frac{\dot{a}}{a}\Bigr)^2 \dual{\lambda} \lambda \,, 
      \label{eq:cos5}
\end{align}
where it was moreover used that
\begin{align}
      \Gamma_x\Gamma_x =\Gamma_y\Gamma_y =\Gamma_z\Gamma_z = 
      \frac{1}{4}\dot{a}^2 \mathbbm{1} \,.
      \label{eq:cos6}
\end{align}
Hence the explicit form of the Lagrangian density is given by
\begin{align}
      \lan_{\rm elko} = \frac{1}{2} \Bigl( \p_t \dual{\lambda} \p_t \lambda
      + \frac{3}{4} \Bigl(\frac{\dot{a}}{a}\Bigr)^2 \dual{\lambda} \lambda
      -m^2 \dual{\lambda} \lambda + \alpha [\dual{\lambda} \lambda]^2 \Bigr) \,,
      \label{eq:cos7}
\end{align}
which should not be varied with respect to the matter fields, see the
cautionary note above.

Inserting the later form into~(\ref{eq:cos4}) yields explicitely the  
components of the energy-momentum tensor
\begin{align}
      T_{tt} = \nabla_{t} \dual{\lambda} \nabla_{t} \lambda - \lan_{\rm elko} 
      = \p_t \dual{\lambda} \p_t \lambda - \lan_{\rm elko} \,,
      \label{eq:cos8}
\end{align}
where $\Gamma_t = 0$ was used, i.e.~$\nabla_t \lambda = \p_t \lambda$, so that
\begin{align}
      T_{tt} = \frac{1}{2} \Bigl( \p_t \dual{\lambda} \p_t \lambda
      - \frac{3}{4} \Bigl(\frac{\dot{a}}{a}\Bigr)^2 \dual{\lambda} \lambda
      + m^2 \dual{\lambda} \lambda - \alpha [\dual{\lambda} \lambda]^2 \Bigr) \,.
      \label{eq:cos9}
\end{align}
Similarly, for the three equal ($T_{xx}=T_{yy}=T_{zz}$) spatial components of the 
energy-momentum tensor one finds
\begin{align}
      T_{xx} &= \nabla_{x} \dual{\lambda} \nabla_{x} \lambda + a^2 \lan_{\rm elko} 
      = -\dual{\lambda}\Gamma_x \Gamma_x \lambda + a^2 \lan_{\rm elko} 
      \nonumber \\
      &= -\frac{1}{4}\dot{a}^2 \dual{\lambda} \lambda + \frac{a^2}{2}
      \Bigl( \p_t \dual{\lambda} \p_t \lambda + \frac{3}{4} 
      \Bigl(\frac{\dot{a}}{a}\Bigr)^2 \dual{\lambda} \lambda
      -m^2 \dual{\lambda} \lambda + \alpha [\dual{\lambda} \lambda]^2 \Bigr) \,,
      \label{eq:cos9a}
\end{align}
which finally yields the following expression
\begin{align}
      T_{xx} = \frac{a^2}{2} \Bigl( \p_t \dual{\lambda} \p_t \lambda + \frac{1}{4} 
      \Bigl(\frac{\dot{a}}{a}\Bigr)^2 \dual{\lambda} \lambda
      -m^2 \dual{\lambda} \lambda + \alpha [\dual{\lambda} \lambda]^2 \Bigr) \,.
      \label{eq:cos10}
\end{align}
One can easily verify that the off-diagonal spatial components of the
energy-momentum tensor for Elko spinors identically vanish. However, the
time-space components $T_{tx}$, $T_{ty}$ and $T_{tz}$, which would
correspond to heat fluxes, require a careful analysis.

Obviously, the terms $g_{tx}\lan_{\rm elko}\,,\ldots$ are absent in the 
energy-momentum tensor because the metric is symmetric. For the component
$T_{tx}$ the remaining term reads
\begin{align}
      T_{tx} = \frac{1}{2}\bigl(\nabla_{t} \dual{\lambda} \nabla_{x} \lambda+
      \nabla_{x} \dual{\lambda} \nabla_{t} \lambda \bigr) =
      \frac{1}{2}\bigl(\p_{t} \dual{\lambda} (-\Gamma_{x} \lambda) +
      \dual{\lambda} \Gamma_{x} \p_{t} \lambda \bigr) \,,
      \label{eq:cos11}
\end{align} 
which with~(\ref{eq:cos3}) can be written as 
\begin{align}
      T_{tx} = \frac{1}{2}\bigl(-\dot{a} \p_{t} \dual{\lambda} f^{01} \lambda + 
      \dot{a} \dual{\lambda} f^{01} \p_{t} \lambda \bigr) \,,
      \qquad f^{01} = \frac{1}{2}\begin{pmatrix} \sigma^1 & \mathbb{O} \\ 
                      \mathbb{O} & -\sigma^1 \end{pmatrix} \,.
      \label{eq:cos12}
\end{align}
Accordingly for the remaining two components are
\begin{align}
      T_{ty} &= \frac{1}{2}\bigl(-\dot{a} \p_{t} \dual{\lambda} f^{01} \lambda + 
      \dot{a} \dual{\lambda} f^{01} \p_{t} \lambda \bigr) \,,
      \qquad f^{02} = \frac{1}{2}\begin{pmatrix} \sigma^2 & \mathbb{O} \\ 
                      \mathbb{O} & -\sigma^2 \end{pmatrix} \,.
      \label{eq:cos13} \\
      T_{tz} &= \frac{1}{2}\bigl(-\dot{a} \p_{t} \dual{\lambda} f^{01} \lambda + 
      \dot{a} \dual{\lambda} f^{01} \p_{t} \lambda \bigr) \,,
      \qquad f^{03} = \frac{1}{2}\begin{pmatrix} \sigma^3 & \mathbb{O} \\ 
                      \mathbb{O} & -\sigma^3 \end{pmatrix} \,.
      \label{eq:cos14}
\end{align}
These three components of the energy-momentum tensor do not vanish
on general grounds for a generic spinor $\lambda$. In order to get
a self-consistent set of equations of motion we must therefore read
the vanishing of equations~(\ref{eq:cos12})--(\ref{eq:cos14}) as 
matter restrictions, i.e.~on the spinors. This is of course expected, since
the spacetime under consideration is homogeneous and isotropic. This
requirement reduces the the metric from 10 independent to just
one function. Hence it is expected, that also the number of 
independent functions for the spinors should reduce. 

Before the consequences of these matter restrictions are discussed, let us
shortly mention the matter equations of motion. Variation of the 
action~(\ref{eq:d14}) with respect to the spinors 
$\dual{\lambda}$ and $\lambda$ respectively yields the equations 
of motion
\begin{align}
      \Box \lambda + m^2 \lambda + 2\alpha [\dual{\lambda}\lambda ] \lambda = 0 \,,
      \label{eq:cos8q} \\
      \Box \dual{\lambda} + m^2 \dual{\lambda} + 2\alpha [\dual{\lambda}\lambda ] 
      \dual{\lambda} = 0 \,,
      \label{eq:cos9q}
\end{align}
where $\Box\lambda = g^{\mu\nu}\nabla_{\mu}\nabla_{\nu}\lambda$ for
the FRW metric~(\ref{eq:cos1a}) is given by 
\begin{align}
      \Box\lambda = \p_{tt} \lambda + 3\Bigl(\frac{\dot{a}}{a}\Bigr)\p_t \lambda -
      \frac{3}{4} \Bigl(\frac{\dot{a}}{a}\Bigr)^2 \lambda \,.
      \label{eq:cos10q}
\end{align}
Note the difference between $\Box\lambda$ and the scalar field analogue
$\Box\phi$ in equation~(\ref{eq:cs7}).

\subsection{Elko's matter restrictions}

For self-consistency of the coupled Einstein-Elko system one must require
\begin{align}
      T_{tx} \stackrel{!}{=} 0 \,, \qquad
      T_{ty} \stackrel{!}{=} 0 \,, \qquad
      T_{tz} \stackrel{!}{=} 0 \,,
      \label{eq:cos14a}
\end{align}
which puts restrictions on the Elko spinors that we will work
out explicitely now. The vanishing of these components is also
a consequence of the field equations, since the geometric part
of them is diagonal, i.e.~$G_{tx}=G_{ty}=G_{tz}=0$. However, the 
notion of matter restrictions might be slightly more appropriate, 
since before coupling any matter, the matter should obey the same 
symmetry properties as the spacetime. However, the approach followed 
here seems to be more direct and better accessible. 

Following~(\ref{eq:d9}), specify the left-handed spinors $\phi_L^{+}$
and $\phi_L^{-}$ with respective helicity up and down to be
\begin{align}
      \phi_L^{+} = \begin{pmatrix} A \\ B \end{pmatrix} \,, \qquad
      \phi_L^{-} = \begin{pmatrix} P \\ Q \end{pmatrix} \,,
      \label{eq:cos15}
\end{align}
where $A=A(t)$, $B=B(t)$, $P=P(t)$ and $Q=Q(t)$ are complex valued 
functions of the time variable $t$. With these definitions the Elko
spinors~(\ref{eq:d9}) become
\begin{align}
      \lambda_{\{-,+\}} = 
      \begin{pmatrix} \mp i \bar{B}\\ \pm i\bar{A}\\ A\\ B\end{pmatrix} \,,
      \qquad
      \lambda_{\{+,-\}} =
      \begin{pmatrix} \mp i \bar{Q}\\ \pm i\bar{P}\\ P\\ Q\end{pmatrix} \,,
      \label{eq:cos16}
\end{align}
where we recall that the upper and lower signs correspond to the
self and anti self conjugate spinors. Sometimes the corresponding spinors 
will be denoted by superscripts $^S$ and $^A$ respectively.
In order to work out these matter restriction, we follow the author's
previous work~\cite{Balasin:2004gf}, where Dirac spinors were 
spherically reduced.

From the spinors~(\ref{eq:cos16}) the following dual spinors by~(\ref{eq:d10})
can be derived
\begin{align}
      \dual{\lambda}_{\{-,+\}} &= +i \lambda_{\{+,-\}}^{\dagger} \gamma^0 =
      \begin{pmatrix} i \bar{P}& i \bar{Q}& \mp Q& \pm P\end{pmatrix} \,,
      \nonumber \\
      \dual{\lambda}_{\{+,-\}} &= -i \lambda_{\{-,+\}}^{\dagger} \gamma^0 =
      \begin{pmatrix} -i \bar{A}& -i \bar{B}& \pm B& \mp A\end{pmatrix} \,.
      \label{eq:cos17}
\end{align}
To avoid calculations involving complex valued functions, henceforth
the functions $A=A(t)$, $B=B(t)$, $P=P(t)$ and $Q=Q(t)$ are written as
\begin{alignat}{2}
      A &= A_0 + i A_1 \,, &\qquad B &= B_0 + i B_1 \,, 
      \nonumber \\
      P &= P_0 + i P_1 \,, &\qquad Q &= Q_0 + i Q_1 \,.
      \label{eq:cos18}
\end{alignat}
Let us assume that the spinor in~(\ref{eq:cos12}) is given by
$\lambda_{\{-,+\}}^S$, the self conjugate Elko spinor with helicities
down and up. This yields the following explicit form of the $(tx)$-component
of the energy-momentum tensor
\begin{multline}
      T_{tx}[\lambda = \lambda_{\{-,+\}}^S] = \frac{\dot{a}}{2} \bigl( 
      A_0 \dot{P}_0 - \dot{A}_0 P_0 - A_1 \dot{P}_1 + \dot{A}_1 P_1\\ -
      B_0 \dot{Q}_0 + \dot{B}_0 Q_0 + B_1 \dot{Q}_1 - \dot{B}_1 Q_1
      \bigr) \,.
      \label{eq:cos19} 
\end{multline}
Had one started with the anti self conjugate spinor $\lambda_{\{-,+\}}^A$
in equation~(\ref{eq:cos12}), one would have obtained the same
result with an overall factor of $-1$, which can easily be verified.
On the other hand, one could have started out the calculation with the
opposite helicity spinors $\lambda_{\{+,-\}}^{S,A}$, which have the
helicity structure up and down. However, the results obtained so,
again only differ by overall factors of $-1$, so that one can summarise
the results for the four possible choices to be
\begin{align}
      T_{tx}[\lambda = \lambda_{\{-,+\}}^S] = -T_{tx}[\lambda = \lambda_{\{-,+\}}^A] =
     -T_{tx}[\lambda = \lambda_{\{+,-\}}^S] = T_{tx}[\lambda = \lambda_{\{+,-\}}^A] \,.
      \label{eq:cos20}
\end{align}
The same of course holds for the two other components of the
energy-momentum tensor, which explicitely read
\begin{multline}
      T_{ty}[\lambda = \lambda_{\{-,+\}}^S] = \frac{\dot{a}}{2} \bigl( 
     -A_0 \dot{P}_1 + \dot{A}_0 P_1 - A_1 \dot{P}_0 + \dot{A}_1 P_0 \\-
      B_0 \dot{Q}_1 + \dot{B}_0 Q_1 - B_1 \dot{Q}_0 + \dot{B}_1 Q_0
      \bigr) \,,
      \label{eq:cos21} 
\end{multline}
and for the last component we obtain
\begin{multline}
      T_{tz}[\lambda = \lambda_{\{-,+\}}^S] = \frac{\dot{a}}{2} \bigl( 
     -A_0 \dot{Q}_0 + \dot{A}_0 Q_0 + A_1 \dot{Q}_1 - \dot{A}_1 Q_1 \\-
      B_0 \dot{P}_0 + \dot{B}_0 P_0 + B_1 \dot{P}_1 - \dot{B}_1 P_1
      \bigr) \,.
      \label{eq:cos22} 
\end{multline}
As already mentioned before, these three components must vanish identically
in order to have a consistent set of field equations. Had we taken any
other spinor of the four possible as the starting point, we would have
only gotten an overall sign, which becomes unimportant if the quantity has
to vanish. Therefore, the same matter restrictions for all four
possible Elko spinors are equal. Taking the consistency requirement~(\ref{eq:cos14a})
into account, the matter has to satisfy the following relations
\begin{subequations}
\begin{align}
      A_0 \dot{P}_0 - \dot{A}_0 P_0 - A_1 \dot{P}_1 + \dot{A}_1 P_1 -
      B_0 \dot{Q}_0 + \dot{B}_0 Q_0 + B_1 \dot{Q}_1 - \dot{B}_1 Q_1 = 0 \,,
      \label{eq:cos23.1}
\end{align}
\begin{align}
     -A_0 \dot{P}_1 + \dot{A}_0 P_1 - A_1 \dot{P}_0 + \dot{A}_1 P_0 -
      B_0 \dot{Q}_1 + \dot{B}_0 Q_1 - B_1 \dot{Q}_0 + \dot{B}_1 Q_0 = 0 \,,
      \label{eq:cos23.2}
\end{align}
\begin{align}
     -A_0 \dot{Q}_0 + \dot{A}_0 Q_0 + A_1 \dot{Q}_1 - \dot{A}_1 Q_1 -
      B_0 \dot{P}_0 + \dot{B}_0 P_0 + B_1 \dot{P}_1 - \dot{B}_1 P_1 = 0 \,.
      \label{eq:cos23.3}
\end{align}
    \label{eq:cos23}
\end{subequations}
According to the rest frame spinors~(\ref{eq:d16}) where $\hat{\vp}=(0,0,1)$ let 
us start with the natural simplifying choice of setting $B=0$ and $P=0$, which 
means that we choose~(\ref{eq:cos15}) of the following form
\begin{align}
      \phi_L^{+} = A \begin{pmatrix} 1 \\ 0 \end{pmatrix} 
      = A\, \mathbf{u}\,, \qquad
      \phi_L^{-} = Q \begin{pmatrix} 0 \\ -1 \end{pmatrix} 
      = Q\, \mathbf{v} \,,
      \label{eq:cos23a}
\end{align}
where the basis vectors were denoted by $\mathbf{u}$ and $\mathbf{v}$.
Note the minus sign in $\phi_L^{-}$, due to the rest frame 
spinors~(\ref{eq:d16}). In that case the equations~(\ref{eq:cos23}) 
simply reduce to
\begin{align}
     A_0 \dot{Q}_0 - \dot{A}_0 Q_0 - A_1 \dot{Q}_1 + \dot{A}_1 Q_1 = 0 \,,
      \label{eq:cos24}
\end{align}
where~(\ref{eq:cos23.1}) and~(\ref{eq:cos23.2}) were automatically satisfied,
and we are left with only one additional condition that has to be satisfied 
by the matter. The next natural choice is to take the two-component building
block of the one spinor real and the other imaginary.

Hence, if one furthermore sets ({\bf a}) $A_1 = 0$ and $Q_0 = 0$ then also 
equation~(\ref{eq:cos24}) is satisfied and one finally has a diagonal
energy-momentum tensor as required by the FRW metric and its resulting
Einstein tensor. Alternatively, one can choose ({\bf b}) $A_0 = 0$ and 
$Q_1 = 0$, which also solves the remaining condition. For the first case 
this corresponds to choosing the spinor $\phi_L^{+}$ real and 
$\phi_L^{-}$ imaginary, and vice versa for the second case
\begin{alignat}{2}
      \mbox{{\bf a:}}\quad  
      \phi_L^{+} &= A_0\, \mathbf{u} \,, &\qquad
      \phi_L^{-} &= i Q_1\, \mathbf{v}\,,
      \label{eq:cos25} \\[1ex]
      \mbox{{\bf b:}}\quad  
      \phi_L^{+} &= i A_1\, \mathbf{u} \,, &\qquad
      \phi_L^{-} &= Q_0\, \mathbf{v} \,.
      \label{eq:cos26}
\end{alignat}
The explicit form of the complete Elko spinor can now be read off
from equations~(\ref{eq:cos16}) and~(\ref{eq:cos17}).

Next, let us compute the norm of the spinors $\dual{\lambda}\lambda$, 
satisfying the three equations~(\ref{eq:cos23}). It turns out that the norm 
vanishes in both cases 
\begin{align}
      \mbox{{\bf a}, {\bf b}:}  
      \quad \dual{\lambda}\lambda = 0\,.
      \label{eq:cos31} 
\end{align}
Not only does the norm of these spinors vanish, furthermore we find
\begin{align}
      \mbox{{\bf a}, {\bf b}:}
      \quad \p_t \dual{\lambda} \p_t \lambda = 0\,.
      \label{eq:cos32} 
\end{align}
This would imply that the energy-momentum tensor for such spinors vanishes
identically, corresponding to ghost Elko spinors. The equations of
motion for the Elko spinors could be solved, however they would not
contribute to the curvature of spacetime. Ghost solutions with vanishing
energy-momentum tensor are quite well known in the context with neutrinos
and they have been studied by several authors in the 1970's, see 
e.g.~\cite{Davis:1974,Griffiths:1979:81}. In the next 
section some analytical ghost Elko solutions will be presented. 

Since one is also interested in non-ghost Elko matter, one should restart 
satisfying the equations~(\ref{eq:cos23}) so that the norm of the spinors 
does not vanish and the matter does contribute to the curvature of spacetime. 
The norm of the spinors reads
\begin{align}
      \dual{\lambda}\lambda[\lambda = \lambda_{\{-,+\}}^S] =
      2 \bigl( - A_0 Q_0 + A_1 Q_1 + B_0 P_0 - B_1 P_1 \bigr) \,,
      \label{eq:cos33}
\end{align}
and similar to equation~(\ref{eq:cos20}) we find
\begin{align}
      \dual{\lambda}\lambda[\lambda = \lambda_{\{-,+\}}^S] =
     -\dual{\lambda}\lambda[\lambda = \lambda_{\{-,+\}}^A] =
     -\dual{\lambda}\lambda[\lambda = \lambda_{\{+,-\}}^S] =
      \dual{\lambda}\lambda[\lambda = \lambda_{\{+,-\}}^A] \,.
      \label{eq:cos34}
\end{align}
In addition to requiring equations~(\ref{eq:cos23}) one adds the
condition of having a non-vanishing norm 
\begin{align}
      \dual{\lambda}\lambda \stackrel{!}{\neq} 0 \,.
      \label{eq:cos34a}
\end{align}
which written out in components leads to the condition
\begin{align}
      - A_0 Q_0 + A_1 Q_1 + B_0 P_0 - B_1 P_1 \neq 0 \,.
      \label{eq:cos35}
\end{align}

Let us start, as before, by the natural choice $B=P=0$ with 
spinors~(\ref{eq:cos23a}) and note again the minus sign that was 
inserted to agree with~(\ref{eq:d15}). Now, we also take the 
non-vanishing of the norm of the spinors into account. 
Equations~(\ref{eq:cos23}) and the norm~(\ref{eq:cos35}) become
\begin{align}
      A_0 \dot{Q}_0 - \dot{A}_0 Q_0 - A_1 \dot{Q}_1 + \dot{A}_1 Q_1 = 0 \,, 
      \label{eq:cos36a} \\
      \dual{\lambda} \lambda = \pm 2 (A_0 Q_0 - A_1 Q_1) \neq 0 \,,
      \label{eq:cos36}
\end{align}
where the upper and lower sign corresponds to the self and anti self
conjugate spinors (see equation~(\ref{eq:d5}) and thereafter).

As was already noticed before, one cannot satisfy both conditions simultaneously
by setting two of the remaining four free functions in~(\ref{eq:cos36}) equal 
to zero. However, one may proceed in the following way. Let us assume that the 
first and the second pair of terms in~(\ref{eq:cos36a}) vanish independently,
so that we arrive at the following three conditions
\begin{align}
      A_0 \dot{Q}_0 - \dot{A}_0 Q_0 = 0 \qquad \Rightarrow \qquad
      Q_0 = \alpha_0 A_0 \,, 
      \nonumber \\ 
      A_1 \dot{Q}_1 - \dot{A}_1 Q_1 = 0 \qquad \Rightarrow \qquad
      Q_1 = \alpha_1 A_1 \,, 
      \nonumber \\
      \dual{\lambda} \lambda = \pm 2 ( A_0 Q_0 + A_1 Q_1 ) = 
      \pm 2 ( \alpha_0 A_0^2 - \alpha_1 A_1^2 ) \neq 0 \,.
      \label{eq:cos37}
\end{align}
In order to simplify the spinors as much as possible we can put
one of the two remaining free functions equal to zero, ({\bf I})
$Q_0 = \alpha_0 A_0 = 0$ or ({\bf II}) $Q_1 = \alpha_1 A_1 = 0$, so 
that
\begin{alignat}{4} 
      \mbox{{\bf I:}}\quad  
      \phi_L^{+} &= i A_1\, \mathbf{u} \,, &\quad
      \phi_L^{-} &= i \alpha_1 A_1\, \mathbf{v} \,, &\qquad
      \dual{\lambda} \lambda &= \pm 2 (- \alpha_1 A_1^2 ) \,,
      \label{eq:cos38} \\[1ex]
      \mbox{{\bf II:}}\quad  
      \phi_L^{+} &= A_0\, \mathbf{u} \,, &\qquad
      \phi_L^{-} &= \alpha_0 A_0\, \mathbf{v} \,, &\qquad
      \dual{\lambda} \lambda &= \pm 2 (\alpha_0 A_0^2 ) \,.
      \label{eq:cos39}
\end{alignat}
Finally we are left with one independent function for both, the self and 
the anti self conjugate Elko spinor. This result is not unexpected,
as will be explained. The cosmological Einstein field equations for the 
FRW metric are just two independent field equations, that imply 
energy-momentum conservation by virtue of the contracted Bianchi identities.
The geometrical part of the field equations contains only one unknown function,
namely the scale factor $a(t)$. Hence, a consistent set a field equations
requires also just one unknown function on the matter side. The above
derived solution to the matter restrictions yields exactly the expected.

Some further remarks are appropriate in order to close this section. It was
natural in the above to assume that the functions $B=P=0$ vanish based on
the zero momentum rest frame spinors~(\ref{eq:d16}). However, this already
assumed a very specific direction of the momentum, namely $\hat{\vp}=\mathbf{z}=(0,0,1)$.
Had one started with another choice, $\hat{\vp}=\mathbf{y}=(0,1,0)$ say, there would
still be one free function, but the purely spinorial parts would have a different
direction. Finally, one could have started with the generic form of the zero
momentum rest frame spinors~(\ref{eq:d15}) to obtain the most general form
of the Elko spinors compatible with the cosmological principle. In any case,
the qualitative results would not change considerably, since there must remain
one free function on the matter side.

One should also shortly comment on the assumption that the two pairs of term
in equation~(\ref{eq:cos36a}) vanish separately. This is also a simplification
that in the generic discussion should not be taken into account. However, it was
the aim of the above construction to find a `minimal' Elko spinor that satisfies
the cosmological principle of homogeneity and isotropy on large scales. This
`minimal' Elko spinor is still rich enough in its structure to possibly account
for an alternative to inflation, as will be shown in the next but one section.

\section{Dynamical aspects of the cosmological Einstein-Elko system}
\label{sec:dyn}

Having solved the conditions that the matter has to satisfy in order
to yield a diagonal energy-momentum tensor, one can now turn to
the cosmological field equations of the Einstein-Elko system.

The gravitational field equations for the FRW metric~(\ref{eq:cos1a})
reduce to two independent equations
\begin{align}
      G_{tt} = 3 \Bigl(\frac{\dot{a}}{a}\Bigr)^2 &= 
      3 H^2 = \kappa \, T_{tt} \,,
      \label{eq:dy1} \\
      G_{xx} = - \dot{a}^2 - 2\,a\,\ddot{a} &= 
      -3 H^2 - 2 \dot{H} = \kappa \, T_{xx}\,,
      \label{eq:dy2} 
\end{align}
where $\kappa$ if the gravitational coupling constant and $H=\dot{a}/a$
denotes the Hubble parameter that in some calculations turns out
to be quite useful. Moreover we took into account that 
$G_{xx}=G_{yy}=G_{zz}$ and also $T_{xx}=T_{yy}=T_{zz}$. 
Differentiation of the first field equation and elimination of
the term $\dot{H}$ yields the conservation equation (the contracted
Bianchi identity)
\begin{align}
      \dot{T}_{tt} + 3 H (T_{tt} + T_{xx}) = 0 \,.
      \label{eq:dy2a}
\end{align}

With~(\ref{eq:cos9}) and~(\ref{eq:cos10}) the field equation explicitely
written out become
\begin{align}
      3 \Bigl(\frac{\dot{a}}{a}\Bigr)^2 = 
      \kappa\,\frac{1}{2} \Bigl( \p_t \dual{\lambda} \p_t \lambda
      - \frac{3}{4} \Bigl(\frac{\dot{a}}{a}\Bigr)^2 \dual{\lambda} \lambda
      + m^2 \dual{\lambda} \lambda - \alpha [\dual{\lambda} \lambda]^2 \Bigr) \,,
      \label{eq:dy3} \\
      -\Bigl(\frac{\dot{a}}{a}\Bigr)^2 - 2\frac{\ddot{a}}{a} = 
      \kappa\,\frac{1}{2} \Bigl( \p_t \dual{\lambda} \p_t \lambda + \frac{1}{4} 
      \Bigl(\frac{\dot{a}}{a}\Bigr)^2 \dual{\lambda} \lambda
      -m^2 \dual{\lambda} \lambda + \alpha [\dual{\lambda} \lambda]^2 \Bigr) \,.
      \label{eq:dy4}
\end{align}
The energy conservation equation can also be made explicit by using~(\ref{eq:dy2a})
together with~(\ref{eq:cos9}) and~(\ref{eq:cos10}), however, this lengthy equation
is suppressed since it does not contain new information. The matter fields also satisfy 
its corresponding equations of motion, given by~(\ref{eq:cos8q}) and the dual 
equation~(\ref{eq:cos9q}), which are not independent due to the conservation
equation.  

The above system of differential equations~(\ref{eq:dy3})--(\ref{eq:dy4}) can
be studied analytically or numerically with suitable initial conditions. If one
assumes the Elko matter to be ghost like, i.e.~with vanishing energy-momentum
tensor, then one can easily solve these equations analytically. In the non-ghost
case it seems quite hopeless to find analytical solutions since the system is 
highly non-linear.

As was already pointed out in section~\ref{ssec:elko}, if we choose the 
spinors to satisfy the matter constraints as put forward in the cases 
{\bf a} or {\bf b}, then the energy-momentum tensor vanishes identically.
Taking into account $\dual{\lambda} \lambda =0$, the field equations and 
matter equations of motion simplify to the following set of equations
\begin{align}
      3 \Bigl(\frac{\dot{a}}{a}\Bigr)^2 = 0 \,, \qquad
      -\dot{a}^2 - 2\,a\,\ddot{a} = 0 \,, 
      \label{eq:dy7} \\[1ex]
      \p_{tt}\lambda + 3\Bigl(\frac{\dot{a}}{a}\Bigr) \p_t \lambda
      - \frac{3}{4} \Bigl(\frac{\dot{a}}{a}\Bigr)^2 \lambda
      + m^2 \lambda = 0 \,,
      \label{eq:dy8} \\
      \p_{tt}\dual{\lambda} + 3\Bigl(\frac{\dot{a}}{a}\Bigr) \p_t \dual{\lambda} 
      - \frac{3}{4} \Bigl(\frac{\dot{a}}{a}\Bigr)^2 \dual{\lambda}
      + m^2 \dual{\lambda} = 0 \,,
      \label{eq:dy9}
\end{align}
Equations~(\ref{eq:dy7}) can easily be solved and we find as a general
solution that the expansion parameter $a(t)=a_g={\rm const.}$ must be 
some constant. Taking this into account, the matter equations of motion 
reduce to 
\begin{align}
      \p_{tt}\lambda + m^2 \lambda = 0 \,,
      \nonumber \\
      \p_{tt}\dual{\lambda} + m^2 \dual{\lambda} = 0 \,,
      \label{eq:dy11}
\end{align}
which is simply the differential equation for the harmonic oscillator,
which is independent of the precise spinor structure of the Elkos.
Nevertheless, for sake of completeness, let us assume that we have the
spinors of case {\bf a} where the complete spinor reads
\begin{align}
      \lambda_{\{-,+\}} = 
      A_0 \begin{pmatrix} 0\\ \pm i\\ 1\\ 0\end{pmatrix} \,,
      \qquad \lambda_{\{+,-\}} =
      i Q_1 \begin{pmatrix} \mp i\\ 0\\ 0\\ -1\end{pmatrix} \,,
      \label{eq:dy12}
\end{align}
with their respective dual spinors given by
\begin{align}
      \dual{\lambda}_{\{-,+\}} &=
      i Q_1 \begin{pmatrix} 0& i& \pm 1& 0\end{pmatrix} \,,
      \nonumber \\
      \dual{\lambda}_{\{+,-\}} &=
      A_0 \begin{pmatrix} -i& 0& 0& \mp 1\end{pmatrix} \,.
      \label{eq:dy13}
\end{align}
Note that it is now irrelevant, which of the four spinors ($\lambda_{\{-,+\}}^S$,
$\lambda_{\{-,+\}}^A$, $\lambda_{\{+,-\}}^S$ or $\lambda_{\{+,-\}}^A$) we use
as our starting point in the equations of motion~(\ref{eq:dy11}), the $\pm$
signs, due to the self or anti self conjugate spinor, drop out of the equations.
In any of the four possible cases the two resulting equations simply become
\begin{align}
      \p_{tt} A_0 + m^2 A_0 = 0 \,,
      \nonumber \\
      \p_{tt} Q_1 + m^2 Q_1 = 0 \,,
      \label{eq:dy14}
\end{align}
which have the general solution
\begin{align}
      A_0(t) = c_1 \sin(mt) + c_2 \cos(mt) \,,
      \nonumber \\
      Q_1(t) = c_3 \sin(mt) + c_4 \cos(mt) \,,
      \label{eq:dy15}
\end{align}
where $c_i$ are arbitrary constants. It is easily verified that these
spinors indeed yield a vanishing norm.

By adding the cosmological constant in the field equations~(\ref{eq:dy7})
the structure of the resulting equations gets slightly more complicated,
however an analytic solution can still be presented. With $\Lambda$ one
has
\begin{align}
      3 \Bigl(\frac{\dot{a}}{a}\Bigr)^2 - \Lambda = 0 \,, \qquad
      -\dot{a}^2 - 2\,a\,\ddot{a} + a^2 \Lambda= 0 \,, 
      \label{eq:dy16}
\end{align}
for which the well known general solution reads as follows
\begin{align}
      a(t) = \exp\Bigl(\pm\sqrt{\frac{\Lambda}{3}}t\Bigr) \,, \qquad
      H(t) = \frac{\dot{a}}{a} = \pm \sqrt{\frac{\Lambda}{3}} \,.
      \label{eq:dy17}
\end{align}
With the cosmological term, the matter equations of motion become
\begin{align}
      \p_{tt}\lambda \pm 3 \sqrt{\frac{\Lambda}{3}} \p_t \lambda
      - \frac{3}{4} \frac{\Lambda}{3} \lambda + m^2 \lambda = 0 \,,
      \label{eq:dy18} \\
      \p_{tt}\dual{\lambda} \pm 3 \sqrt{\frac{\Lambda}{3}} \p_t \dual{\lambda} 
      - \frac{3}{4} \frac{\Lambda}{3} \dual{\lambda}
      + m^2 \dual{\lambda} = 0 \,.
      \label{eq:dy19}
\end{align}
Inserting, as before, the equations~(\ref{eq:dy12}) and~(\ref{eq:dy13}) yields
\begin{align}
      \p_{tt} A_0 + \sqrt{3 \Lambda}\, \p_t A_0
      - \frac{\Lambda}{4} A_0 + m^2 A_0 = 0 \,,
      \label{eq:dy20} \\
      \p_{tt} Q_1 + \sqrt{3 \Lambda}\, \p_t Q_1 
      - \frac{\Lambda}{4} Q_1 + m^2 Q_1 = 0 \,,
      \label{eq:dy21}
\end{align}
where we only took the positive sign into account. The general solution
of the matter equations read
\begin{align}
      A_0(t) &= c_5 \exp(t\tau_{+}/2) + c_6 \exp(t\tau_{-}/2) \,,
      \nonumber \\
      Q_1(t) &= c_7 \exp(t\tau_{+}/2) + c_8 \exp(t\tau_{-}/2) \,,
      \nonumber \\
      \tau_{\pm} &= -\sqrt{3\Lambda} \pm 2\sqrt{\Lambda -m^2} \,.
      \label{eq:dy22}
\end{align}
The negative sign in~(\ref{eq:dy19}) yields the same formal structure
of the solution, but with $\tau_{\pm}=\sqrt{3\Lambda} \pm 2\sqrt{\Lambda -m^2}$. 
As before, it is easy to verify that the norm of the these
spinors vanishes and so does its energy-momentum tensor.
It is quite interesting in this setup that the mass of the spinors
and the cosmological constant are related, compare e.g.~\cite{Dimakis:1985jb},
where ghost solutions of the Einstein-Cartan-Dirac system were analysed and
it was found that the cosmological constant and the mass of the Dirac
fermions are related by $\Lambda=-3m^2$.

The functions $A_0$ and $Q_1$ are real valued by construction, therefore,
in order to have well defined functions in~(\ref{eq:dy22}) we must require
\begin{align}
      m^2 \leq \Lambda \,,
      \label{eq:dy23}
\end{align}
which means that ghost Elko spinors can only exist for relatively small
masses. It is interesting to note that the massless limit in~(\ref{eq:dy22})
is well defined, although the underlying field theory of massless Elko
spinors is subtle and in fact the massless limit of an Elko quantum
field theory is singular (see Section~8.2 of Ref~\cite{jcap}), because
of the theory's non-locality.

\section{Can Elko dark matter drive inflation?}
\label{inflation}

The theory of phase transitions is used to study cosmological applications of 
field theories. The simplest inflationary models are based on a scalar field 
theory with self-interaction term. Inflationary models seem to be an
unavoidable part of cosmology if one wishes to solve the flatness and horizon 
problems~\cite{Guth:1980zm}. Another attractive feature of inflationary models 
is that supersymmetric theories gain strong support from inflation, or as other 
authors put it ``inflation cries out for supersymmetry''~\cite{Dominguez-Tenreiro:1988fz}. 
Since all observational data~\cite{Riess:1998cb,Perlmutter:1998} 
favours flat constant time hypersurface $(k=0)$, this gives strong
support for an inflationary phase in the universe. 

For sake of concreteness, we specialise to the non-ghost case {\bf II}
in which the spinors take the following form
\begin{align}
      \lambda_{\{-,+\}} = 
      A_0 \begin{pmatrix} 0\\ \pm i\\ 1\\ 0\end{pmatrix} \,,
      \qquad \lambda_{\{+,-\}} =
      \alpha_0 A_0 \begin{pmatrix} \pm i\\ 0\\ 0\\ -1\end{pmatrix} \,,
      \label{eq:dy24}
\end{align}
with their respective dual spinors given by
\begin{align}
      \dual{\lambda}_{\{-,+\}} &=
      \alpha_0 A_0 \begin{pmatrix} 0& -i& \pm 1& 0\end{pmatrix} \,,
      \nonumber \\
      \dual{\lambda}_{\{+,-\}} &=
      A_0 \begin{pmatrix} -i& 0& 0& \mp 1\end{pmatrix} \,.
      \label{eq:dy25}
\end{align}
The following norms for the spinors are implied
\begin{align}
      \dual{\lambda}_{\{-,+\}} \lambda_{\{-,+\}} = 
      \dual{\lambda}_{\{+,-\}} \lambda_{\{+,-\}} = 
      \pm 2 \alpha_0 A_0^2 \,.
      \label{eq:dy26}
\end{align}
Let us now put these simplified spinors back into the Elko Lagrangian
\begin{align}
      \lan = \frac{1}{2} \Bigl( g^{\mu\nu}\nabla_{\mu} \dual{\lambda} 
      \nabla_{\nu} \lambda  -m^2 \dual{\lambda} \lambda + \alpha 
      [\dual{\lambda} \lambda]^2 \Bigr)\,,
      \label{eq:dy27}
\end{align}
and analyse its formal structure. At this stage we can use either one of
the two spinors $\lambda_{\{-,+\}}$ or $\lambda_{\{+,-\}}$ and the final
result will be unaffected by that. The three respective terms in the
above Lagrangian explicitely written out become
\begin{align}
      g^{\mu\nu}\nabla_{\mu} \dual{\lambda} \nabla_{\nu} \lambda &=
      \pm 2 \alpha_0 \dot{A}_0^2 \pm \frac{3}{4} 
      \Bigl(\frac{\dot{a}}{a}\Bigr)^2\, 2 \alpha_0 A_0^2 \,,
      \nonumber \\
      m^2 \dual{\lambda} \lambda &= \pm m^2\, 2 \alpha_0 A_0^2 \,,
      \nonumber \\
      \alpha [\dual{\lambda} \lambda]^2 &= \alpha [2 \alpha_0 A_0^2]^2 \,,
      \label{eq:dyn28}
\end{align}
where the $\pm$ drops out in the self-interaction term.
After introducing the new {\em scalar} field $\Phi$ by
\begin{align}
      \Phi = \sqrt{2\alpha_0} A_0\,, \qquad
      \dot{\Phi} = \sqrt{2\alpha_0} \dot{A}_0\,,
      \label{eq:dy29}
\end{align}
the terms~(\ref{eq:dy29}) can be rewritten so that the Elko Lagrangian
in terms of the {\em scalar} field $\Phi$ is given by
\begin{align}
      \lan = \pm \frac{1}{2} \Bigl( \dot{\Phi}^2  + \frac{3}{4} 
      \Bigl(\frac{\dot{a}}{a}\Bigr)^2 \Phi^2 -m^2 \Phi^2 \pm \alpha 
      \Phi^4 \Bigr)\,.
      \label{eq:dy30}
\end{align}
Finally, introduce an effective mass $m_{\rm eff}$ as
\begin{align}
      m_{\rm eff}^2 = m^2 - \frac{3}{4} \Bigl(\frac{\dot{a}}{a}\Bigr)^2 \,,
      \label{eq:dy31}
\end{align}
and let us write the term $\dot{\Phi}^2$ with the help of the covariant
derivative such that $\dot{\Phi}^2 = g^{\mu\nu}\nabla_{\mu}\Phi\nabla_{\nu}\Phi$
(for which it is crucial the $\Phi$ is a scalar field, see equations~(\ref{eq:cs1}) 
and~(\ref{eq:cs3})) the Lagrangian
becomes
\begin{align}
      \lan = \pm \frac{1}{2} \Bigl(  
      g^{\mu\nu}\nabla_{\mu}\Phi\nabla_{\nu}\Phi 
      -m_{\rm eff}^2 \Phi^2 \pm \alpha \Phi^4 \Bigr)\,.
      \label{eq:dy32}
\end{align}
It is this latter (without the effective mass term) scalar field Lagrangian, 
or field theory, with self interaction that gave rise to today's inflationary 
models. 

Elko spinors are prime dark matter candidates since their interaction with
standard model particles is very weak, see~\cite{jcap,prd} for details. The 
simplest cosmological Elko spinors satisfy scalar-like equations of motion 
and allow a (power counting renormalisable) quartic self interaction term. 
Such a model can be mapped to a scalar field theory and it is tempting to 
not only regard the Elko spinors as prime dark matter candidates but also 
to consider them as the actual source of inflation. 

This would put both, dark matter and inflation, on a firm and unified and first 
principle theoretical footing, since the only ingredient is the mass dimension 
one Elko spinor. So, to finally answer the question asked in the title of this 
section, yes, it seems probable that Elko dark matter could drive inflation.

\section{Summary and further outlook}
\label{sec:end}

In the present work the recently discovered Elko spinors were introduced
into spacetime by the standard covariantisation procedure. After assuming 
a cosmological spacetime, analytic ghost Elko solutions were presented.
Finally it was shown that the Elko spinor is a first principle prime
candidate for inflation, having more attractive features than today's models,
because the Elko spinor might account for both, dark matter and inflation.
Since the complete Elko spinor has four independent rest frame spinors,
just like the Dirac spinor, these additional degrees of freedom may also 
account for double inflation~\cite{Silk:1986vc}. This possibility makes
the Elko spinors an even more interesting candidate for the physics of 
inflation.

There are several open questions that are subject to further research.
The theory of phase transitions should be applied to the Elko quantum
field theory, in order to investigate its consequences, in particular 
whether it is indeed a prime candidate for inflation, as suggested here. 
From a conceptual point of view if would be interesting to formulate the 
Elko spinors within the Geroch-Held-Penrose spin coefficient 
formalism~\cite{penrosespinors}. This would further enlighten the geometrical 
structure induced by these spinors. Moreover, along the lines 
of~\cite{Balasin:2004gf}, it would be interesting to study the 
coupled Einstein-Elko system in spherical symmetry.

Another possibility, to further study Elko spinors, is the analysis
of their structure in spacetimes with torsion,
see e.g.~\cite{Hehl:1973:74,Hehl:1976kj}. If the Elko spinors
are minimally coupled, the resulting torsion tensor is quite
complex, for first results see~\cite{Boehmer:2006qq}. This can 
be directly seen from the fact that the Lagrangian is 
non-linear in the spinors and hence also the contortion tensor would
appear quadratically in the matter action. Moreover, terms of
the form $\dual{\lambda}K\nabla\lambda$ are also present in the action,
see~\cite{Hehl:1976kj} for the massive vector field discussion that shows 
similar properties. The sensitivity of Elko spinors for torsion can 
also be understood from their helicity structure. In contrast to the 
Dirac spinors, where both two-spinors have the same helicity, the Elkos 
have a dual helicity structure. It is clear that such an 
opposite helicity configuration is affected by torsion more significantly 
than the aligned helicity structure of Dirac spinors.

Finally I would like to conclude that the Elko spinors provide us with a 
rich source of further investigation, interesting for many aspects of
general relativity and gravitation: inflation, early cosmology and dark 
matter regarding today's problem in standard cosmology; spherical and axial
symmetry for exact solutions; the spin coefficient formalism in order to 
study their geometrical structure; and alternative theories of gravity with 
torsion because of their dual helicity structure.

\acknowledgments
I would very much like to thank Dharam Ahluwalia-Khalilova for making this work possible. 
This project is supported by research grant BO 2530/1-1 of the German 
Research Foundation (DFG).

\end{document}